\documentclass{IOS-Book-Article}
\usepackage{url}
\usepackage{graphicx}
\usepackage[usenames,dvipsnames,svgnames,table]{xcolor}
\usepackage{hyperref}
\usepackage{bchart}

\definecolor{linkblue}{RGB}{0,0,155}

\hypersetup{
   bookmarksnumbered,
   colorlinks={true},
   pdfstartview={FitH},
   citecolor={linkblue},
   linkcolor={linkblue},
   urlcolor={linkblue},
}

\begin{document}
\begin{frontmatter}
\title{Using the AIDA Language to\\Formally Organize Scientific Claims}

\author{\fnms{Tobias} \snm{Kuhn}}
\address{Department of Computer Science, Vrije Universiteit Amsterdam, The~Netherlands}

\begin{abstract}
Scientific communication still mainly relies on natural language written in scientific papers, which makes the described knowledge very difficult to access with automatic means. We can therefore only make limited use of formal knowledge organization methods to support researchers and other interested parties with features such as automatic aggregations, fact checking, consistency checking, question answering, and powerful semantic search. Existing approaches to solve this problem by improving the scientific communication methods have either very restricted coverage, require formal logic skills on the side of the researchers, or depend on unreliable machine learning for the formalization of knowledge.
Here, I propose an approach to this problem that is general, intuitive, and flexible. It is based on a unique kind of controlled natural language, called AIDA, consisting of English sentences that are atomic, independent, declarative, and absolute. Such sentences can then serve as nodes in a network of scientific claims linked to publications, researchers, and domain elements.
I present here some small studies on preliminary applications of this language. The results indicate that it is well accepted by users and provides a good basis for the creation of a knowledge graph of scientific findings.
\end{abstract}

\end{frontmatter}

\section*{Introduction}

It is increasingly difficult for scientists to keep up with the rapidly growing literature, and science is in a genuine communication crisis \cite{bastian2010seventy}. Text mining has been the favorite approach to address this problem, but results remain far from perfect even for basic tasks, such as entity recognition \cite{nadeau2007survey} and extraction of simple relations \cite{wei2015overview}, and such approaches have therefore failed to mimic human capacity to understand texts describing complex results. As an alternative solution, annotation approaches \cite{ciccarese2011open} have been proposed that require humans in the loop to manually attach formal representations and links to existing articles. Such annotations are, however, quite complicated to create, typically apply restrictions on what can be expressed, and can often be understood only in the larger context of the underlying article text.

In earlier work \cite{kuhn2013eswc}, I sketched an alternative approach that goes beyond annotation, is simple and intuitive, is fully general, and leads to representations that can be linked to scientific articles but are completely independent entities. Specifically, this approach is based on the simple idea that we can use single English sentences to capture scientific findings and hypotheses, which then form the nodes in a network of scientific knowledge. My previous work introduced the concept of AIDA sentences \cite{kuhn2013eswc}, which are defined as English sentences that are: \textbf{Atomic:} a sentence describing one thought that cannot be further broken down in a practical way; \textbf{Independent:} a sentence that can stand on its own, without external references like ``this effect'' or ``we''; \textbf{Declarative:} a complete sentence ending with a full stop that could (at least in theory) be true or false; and \textbf{Absolute:} a sentence describing the core of a claim ignoring the (un)certainty about its truth and ignoring how it was discovered (without phrases such as ``probably'' or ``evaluation showed that'').

The language of AIDA sentences thereby forms a Controlled Natural Language (CNL) \cite{kuhn2014cl}. Its approach is based on the assumption that virtually every scientific hypothesis can be represented as such a sentence. In the future, we could ask from researchers to publish their results in such a format from the start, and these AIDA sentences can then serve as nodes in a growing network of scientific claims and as a basis for the formal representation of their content, following our proposed vision of genuine semantic publishing \cite{kuhn2017genuine}.

Some examples of AIDA sentences are shown here:
\begin{itemize}
\item ``A combination of system and searcher biases lead search engine users to settle on the incorrect answer to yes/no-questions around half of the time.'' (from {\small\url{https://doi.org/10.1145/2484028.2484053}})
\item ``Teenagers reply on average faster to emails than adults.'' (from {\small\url{https://doi.org/10.1145/2736277.2741130}})
\item ``Deep learning is a powerful and accurate method for automatic speech recognition.'' (from {\small\url{https://doi.org/10.1109/ASRU.2011.6163930}}, {\small\url{https://doi.org/10.1109/MSP.2012.2205597}}, and {\small\url{https://doi.org/10.1109/ICASSP.2013.6639347}})
\end{itemize}
These examples illustrate the benefits of the different requirements of the AIDA approach. Atomicity ensures that each sentence is as short and concise as possible. Independence allows us to interpret and understand these sentences without further context (the sentences above can be understood without looking at the references we provided). Declarativeness gives them a common form and allows us to categorize them as true or false, or any degree of uncertainty in between. Absoluteness, finally, contributes to normalizing the sentences, thereby allowing us to use the same identifier (i.e. AIDA sentence) for reported results that only differ in their uncertainty or method of discovery, as exemplified by the last sentence above.
These degrees of certainty and these methods of discovery are of course important too, but they are relatively easy to record with classical formal methods and formally linked to AIDA sentences. Various ontologies have in fact been proposed for these aspects (e.g. \cite{dewaard2012formalising} and \cite{chibucos2014standardized}).


These properties make an AIDA sentence highly reusable, and we can treat it as an anchor to formally link, for example, papers that claim or refute it. We can also link the sentences among each other, such as stating that a given AIDA sentence is more specific or more general than another one, or has the same meaning.
Moreover, we can allow for these AIDA sentences themselves to be partially or fully specified in a formal logic language like RDF, thereby allowing for a full continuum from informal over semi-formal to fully formalized statements, as proposed in our earlier work \cite{kuhn2012wole}.

In previous work, we also presented two studies on the manual and automatic creation, respectively, of AIDA sentences in the biomedical field \cite{kuhn2013eswc}. These studies showed that manual creation of AIDA sentences by untrained researchers as well as their automatic creation from an existing biomedical data source can be performed in either case in an effective and accurate manner. In both cases, about 70\% of the created AIDA sentences received a perfect quality score.
In a follow-up study, we worked on the extraction of AIDA sentences from paper abstracts with a simple rule-based approach \cite{jansen2016bnaic}.

AIDA is certainly not the first controlled natural language that aims to improve the way how science is conducted and communicated. In fact, very first English-based CNL, Basic English, was designed around 1930 to improve the global communication in science \cite{ogden1930basic}, among other spheres such as politics and economy. However, Basic English had no relation to formal methods and automatic knowledge organization, but only dealt with inter-human communication.
Formally precise CNLs have been proposed more recently to improve the communication of scientific results \cite{kuhn2006improving} and to provide intuitive yet powerful query interfaces to researchers \cite{hallett2007coli,kuhn2012corpora}. These approaches have however quite narrow application ranges that are limited by the expressiveness of the underlying logic formalism and the coverage of the used vocabularies or ontologies.
AIDA is unique in the sense that it aims to support formal knowledge representation, while focusing on expressiveness rather than precision \cite{kuhn2014cl}.

\section*{Data}

In past few years, I have been building a small dataset of hand-curated AIDA sentences, which serves as the basis for the small studies to be introduced below.
The two studies in the biomedical field introduced above \cite{kuhn2013eswc} provide us with the first batch of AIDA sentences. The manual AIDA extraction study created 51 AIDA sentences and the automatic extraction created another 189 of them, therefore 240 AIDA sentences in total. They all come with an identifier of the associated publication they were extracted from.

\begin{figure}[tb]
\includegraphics[width=\textwidth]{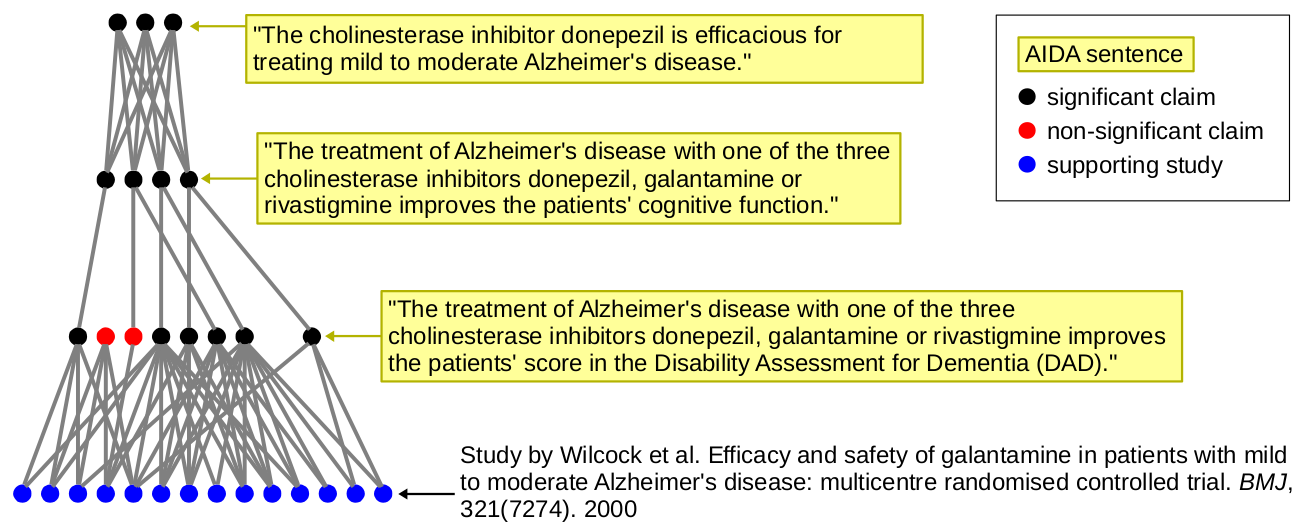}
\caption{Part of the network of AIDA sentences and publications from the Alzheimer's case study. The links between AIDA claims connect a more specific claim (bottom) to a more general one (top).}
\label{fig:alzheimers}
\end{figure}

The next sets of AIDA sentences come from two small case studies on meta-reviews in different domains. The first of these meta-reviews is a Cochrane Library report entitled ``Cholinesterase inhibitors for Alzheimer's disease'' ({\small\url{https://doi.org/10.1002/14651858.CD005593}}), summarizing evidence and findings from a number of publications on the topic. I manually created a network of AIDA sentences and the relevant publications based on the information found in the meta-review report. Figure \ref{fig:alzheimers} shows the main part of the resulting network, with three levels of AIDA sentences (more general ones at the top), and the lowest level being linked to the individual publications. The full data contains 62 AIDA sentences and can be found online.\footnote{\url{https://github.com/tkuhn/aida/blob/master/usecases/alzheimers.md}}

The second case study targeted a less formal meta-review report published by SPARC Europe\footnote{\url{https://sparceurope.org/what-we-do/open-access/sparc-europe-open-access-resources/open-access-citation-advantage-service-oaca/}} on the question of whether Open Access publications enjoy a citation advantage. This general question can be expressed as the AIDA sentence ``Open Access publications receive on average more citations than similar publications that are not Open Access''. Most covered publications, however, investigate a narrower claim such as ``Open Access publications \emph{in astronomy and physics} receive ...''. Overall, this second case study created AIDA sentences for 70 publications (only one paper could not be found and had to be excluded). The details of this study can be found online as well.\footnote{\url{https://github.com/tkuhn/aida/blob/master/usecases/openaccess.md}}

Finally, I have been building a personal collection of AIDA sentences for some of the scientific publications I have read. This collection consists at the moment of 287 AIDA sentences from a wide variety of scientific disciplines. The examples shown in the introduction of this paper are from this collection.

The combined collection therefore consists at the moment of 659 AIDA sentences (650 at the time the network study to be described below was conducted; I have added nine entries to my personal collection since). All these AIDA sentences were manually curated. Automatically extracted sentences were only added after a manual check for accuracy and AIDA compliance.

\section*{User Study}

I felt that AIDA sentences could turn out to be a beneficial technology also in the classroom setting.
I have therefore used AIDA sentences for a Master course that I have been giving at VU Amsterdam during the fall semesters since 2015. In this course, entitled ``Knowledge and Media'', students are required to read 20 given papers on topics around knowledge organization. In order to help them remember and understand what they have read, I provided them with AIDA sentences for each of these 20 publications and for some additional publications that I mentioned in the lecture slides. These sentences are also part of the personal AIDA collection introduced above.

\begin{figure}[tb]
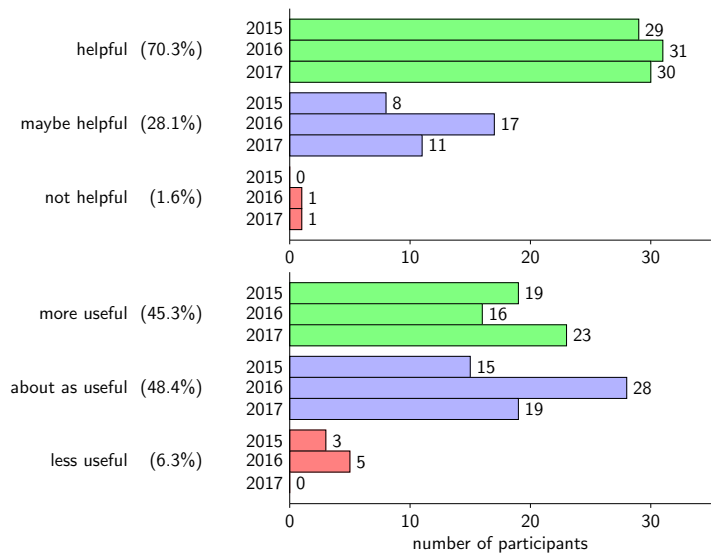

\begin{center}%
\scalebox{0.7}{\begin{bchart}[max=35,step=10,width=10cm,scale=0.8]
\bcbar[label=\phantom{about as useful ~(48.4\%) ~ ~ ~ }2015,color=green!50]{29}
\bcbar[label={helpful ~(70.3\%) ~ ~ ~ 2016},color=green!50]{31}
\bcbar[label=2017,color=green!50]{30}
\smallskip
\bcbar[label=2015,color=blue!30]{8}
\bcbar[label={maybe helpful ~(28.1\%) ~ ~ ~ 2016},color=blue!30]{17}
\bcbar[label=2017,color=blue!30]{11}
\smallskip
\bcbar[label=2015,color=red!50]{0}
\bcbar[label={not helpful ~\phantom{0}(1.6\%) ~ ~ ~ 2016},color=red!50]{1}
\bcbar[label=2017,color=red!50]{1}
\end{bchart}}%
\par%
\scalebox{0.7}{\begin{bchart}[max=35,step=10,width=10cm,scale=0.8]
\bcbar[label=2015,color=green!50]{19}
\bcbar[label={more useful ~(45.3\%) ~ ~ ~ 2016},color=green!50]{16}
\bcbar[label=2017,color=green!50]{23}
\smallskip
\bcbar[label=2015,color=blue!30]{15}
\bcbar[label={about as useful ~(48.4\%) ~ ~ ~ 2016},color=blue!30]{28}
\bcbar[label=2017,color=blue!30]{19}
\smallskip
\bcbar[label=2015,color=red!50]{3}
\bcbar[label={less useful ~\phantom{0}(6.3\%) ~ ~ ~ 2016},color=red!50]{5}
\bcbar[label=2017,color=red!50]{0}
\bcxlabel{number of participants}
\end{bchart}}%
\end{center}
\caption{Responses from the participants on whether AIDA sentences were helpful (top) and on how AIDA sentences compared to classical text summaries (bottom)}
\label{fig:aidaresults}
\end{figure}

In the end of the course, I asked the students to give me feedback on their opinions with respect to AIDA sentences. Specifically, I asked them the following questions with the shown answer options:
\begin{quote}
\textbf{1. AIDA Sentences:} Were the AIDA sentences, as presented during the lectures and on the slides, helpful for you to understand and remember the content of the papers?
\begin{itemize}
\item Yes, the AIDA sentences were helpful.
\item Maybe. I am not sure whether the AIDA sentences were helpful.
\item No, the AIDA sentences were not helpful.
\end{itemize}
\end{quote}
\begin{quote}
\textbf{2. AIDA sentences compared to classical text summaries:} Did you find the AIDA sentences, as presented during the lectures and on the slides, to be more or less useful than classical text summaries?
\begin{itemize}
\item I found the AIDA sentences to be more useful than classical text summaries.
\item I found the AIDA sentences to be about as useful as classical text summaries.
\item I found the AIDA sentences to be less useful than classical text summaries.
\end{itemize}
\end{quote}
In total, 128 students participated over the three years since 2015. The exact set of AIDA sentences varied slightly between the years as I tried to optimize the mix of papers and removed unpopular ones from the list. The students' responses are shown in Figure \ref{fig:aidaresults}.

We see that 70.3\% of all participating students thought the AIDA sentences were helpful to understand and remember the content of the papers. There is some variation over the years, but the \emph{helpful} group formed a clear majority in every single year, and over the three years only two students (out of 128) responded with \emph{not helpful}. If we try to boil this down to a single number by assigning positive answers the value $+1$, \emph{maybe} answers the value $0$, and negative answers the value $-1$, we get an average response of $+0.69$, which is far in the positive range.

These results indicate that AIDA sentences are indeed helpful to a certain extent, but they do not tell us whether this extent is large or small. The answers of the students to the second question, comparing AIDA sentences to classical text summaries, can give us some insights on this. The positive answers (i.e. that AIDA sentences are more useful) still form the majority in the years 2015 and 2017, but not in 2016 and the overall dataset, at 45.3\% of the overall responses. The majority in 2016 as well as the overall majority responded with the neutral answer (i.e. that AIDA sentences are about as useful as classical summaries). On the other hand, only eight out of the 128 respondents would have preferred classical text summaries. If we quantify the overall effect again in a single number in the same manner as before, we get a value of $+0.39$, which is still clearly in the positive range, even though considerably less so than for the first question.

These results indicate that AIDA sentences are indeed an intuitive and accurate method to structure scientific findings. Importantly, a vast majority of students found the AIDA sentences to be not less useful than classical summaries, despite the fact that they did not even get to experience some of the core benefits of AIDA sentences, namely advanced knowledge access powered by the formal interlinking features.

\section*{Linking and Network Study}

While AIDA sentences do not require us to formally represent their domain-level content, they can serve as a tool and structure to support this process. The benefit of the AIDA approach is that this formalization does not have to happen right away (or at all), can be done by a different person (or algorithm) at a later point in time, and allows for any degree of partial formalization. And in the meantime, formal links on the meta-level of statements can be established.
The study presented in this section investigates to what extent a simple kind of such post-hoc partial formalization and linking can lead to a connected knowledge graph of scientific findings.

The aim of this study is to automatically link the AIDA sentences from the datasets above to the Linked Open Data cloud \cite{bizer2009linked}. For that, I applied DBpedia Spotlight \cite{mendes2011dbpedia}, which is an annotation tool automatically linking text in natural language to DBpedia \cite{auer2007dbpedia} identifiers (which map to Wikipedia pages). I used the DBpedia Spotlight API\footnote{\url{http://www.dbpedia-spotlight.org/api}}, with the default confidence parameter of $0.5$. This annotation process of the 650 AIDA sentences of our combined dataset led to overall 1726 DBpedia mappings, i.e. on average 2.66 per AIDA sentence.

In order to assess the quality of these links, I first performed a manual evaluation of a random sample of 10\% of the resulting annotations (173 out of 1726). The result showed that these annotations were correct in 94.2\% of the cases, i.e. a bit more than one error per twenty annotations. The errors range from minor differences, such as annotating the word ``Japanese'' in ``Japanese population'' with the DBpedia term for ``Japanese language'', to completely unrelated concepts, such as mapping the mention of the gene identifier ``CHES-1'' to the DBpedia entry for the ``Workshop on Cryptographic Hardware and Embedded Systems'', which happens to share the same acronym.

The correct annotations also show a broad range, with general and straightforward annotations on the one end, such as ``probability'' in this example from the Alzheimer's study:
\begin{quote}
The treatment of Alzheimer's disease with one of the three cholinesterase inhibitors donepezil, galantamine or rivastigmine has a higher probability of at least one adverse event of anorexia before the end of the treatment as compared to a placebo treatment.
\end{quote}
On the more specific and more complex end, the phrase ``cholinesterase inhibitors'' of the example above is correctly mapped to the DBpedia term ``Acetylcholinesterase inhibitor'', which is a slightly different name for the same thing. Moreover, also ``donepezil'', ``galantamine'', ``rivastigmine'', ``adverse event'', ``anorexia'', and ``placebo'' in the example above are correctly mapped to their DBpedia identifiers.

\begin{figure}[tb]
\includegraphics[width=0.32\textwidth]{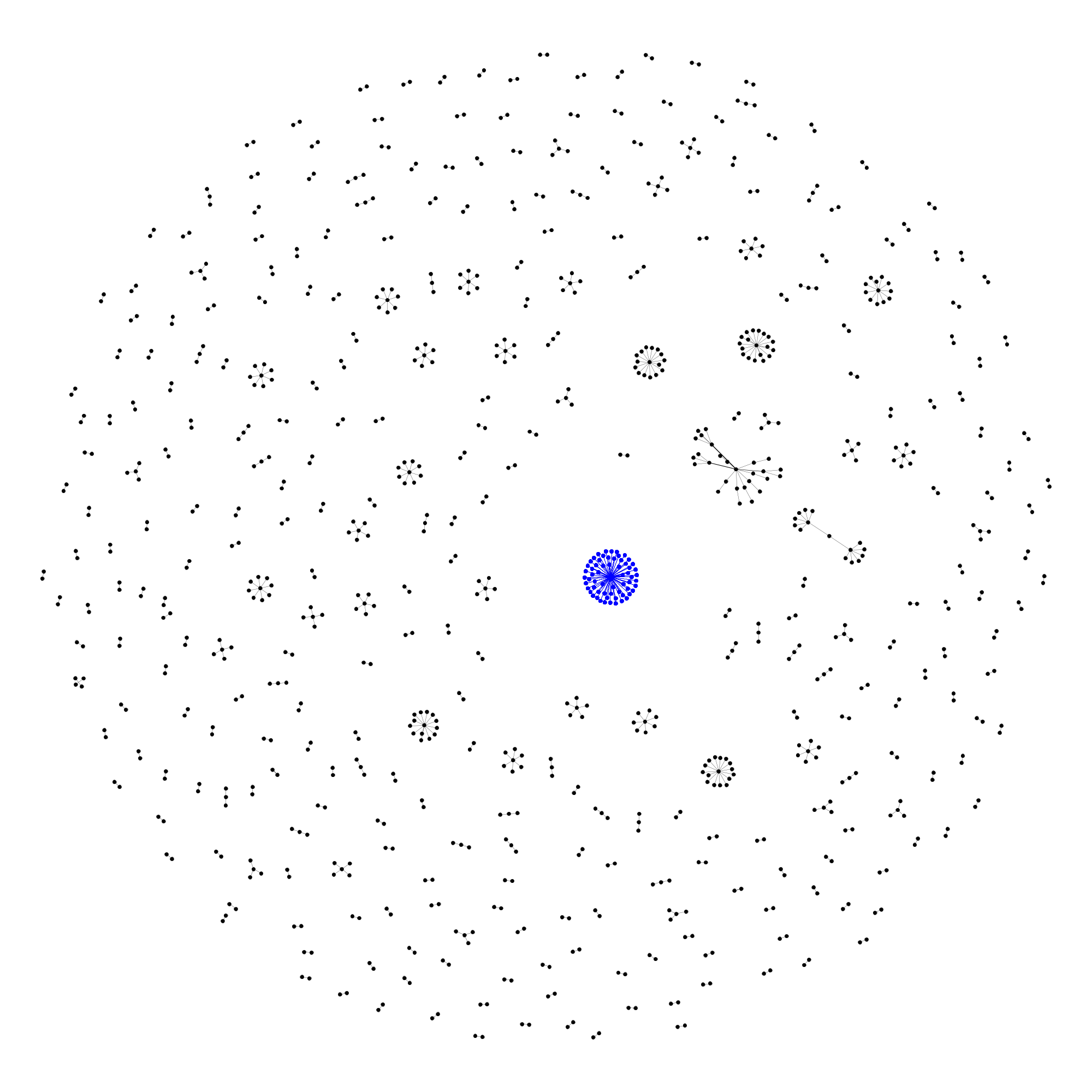}%
\includegraphics[width=0.32\textwidth]{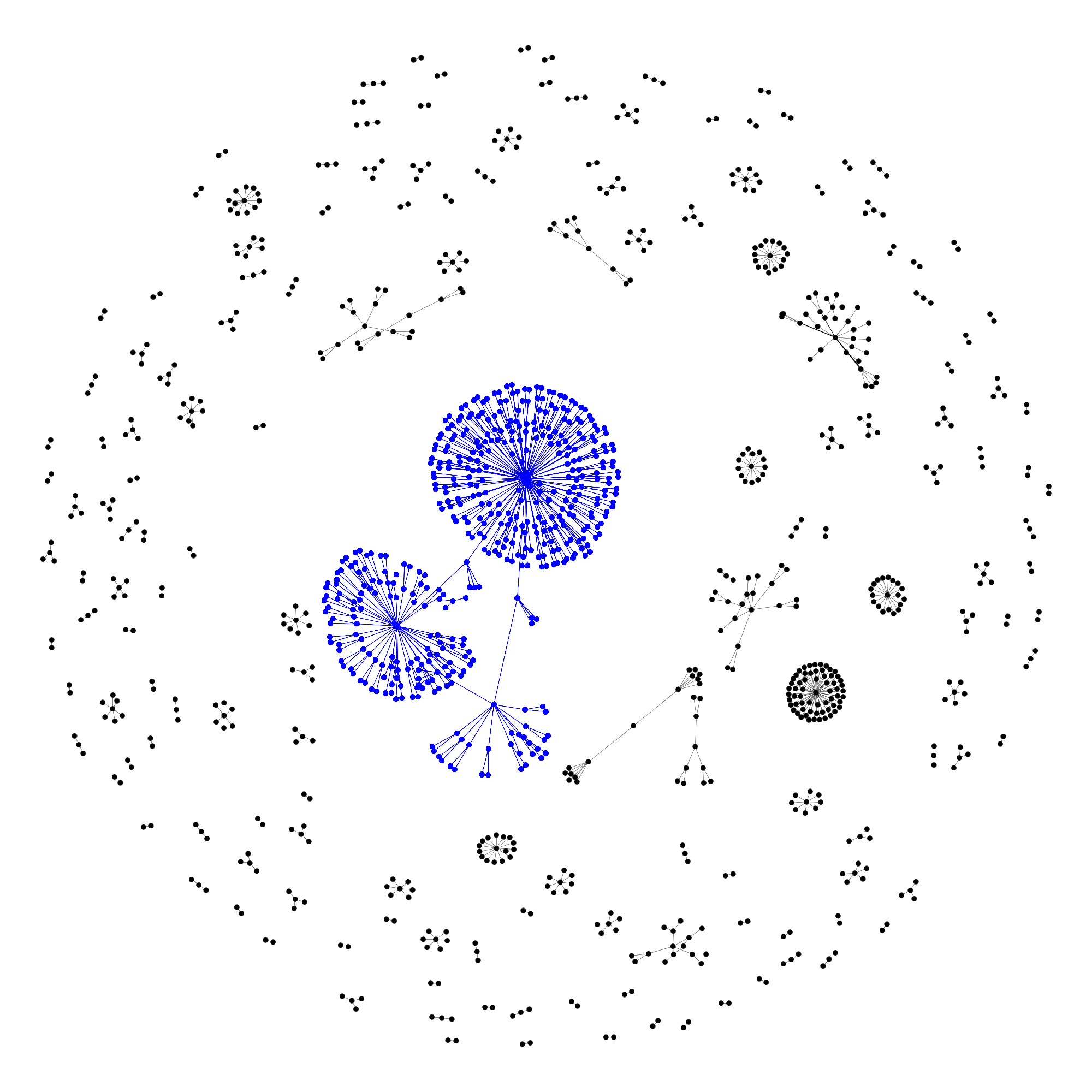}%
\includegraphics[width=0.32\textwidth]{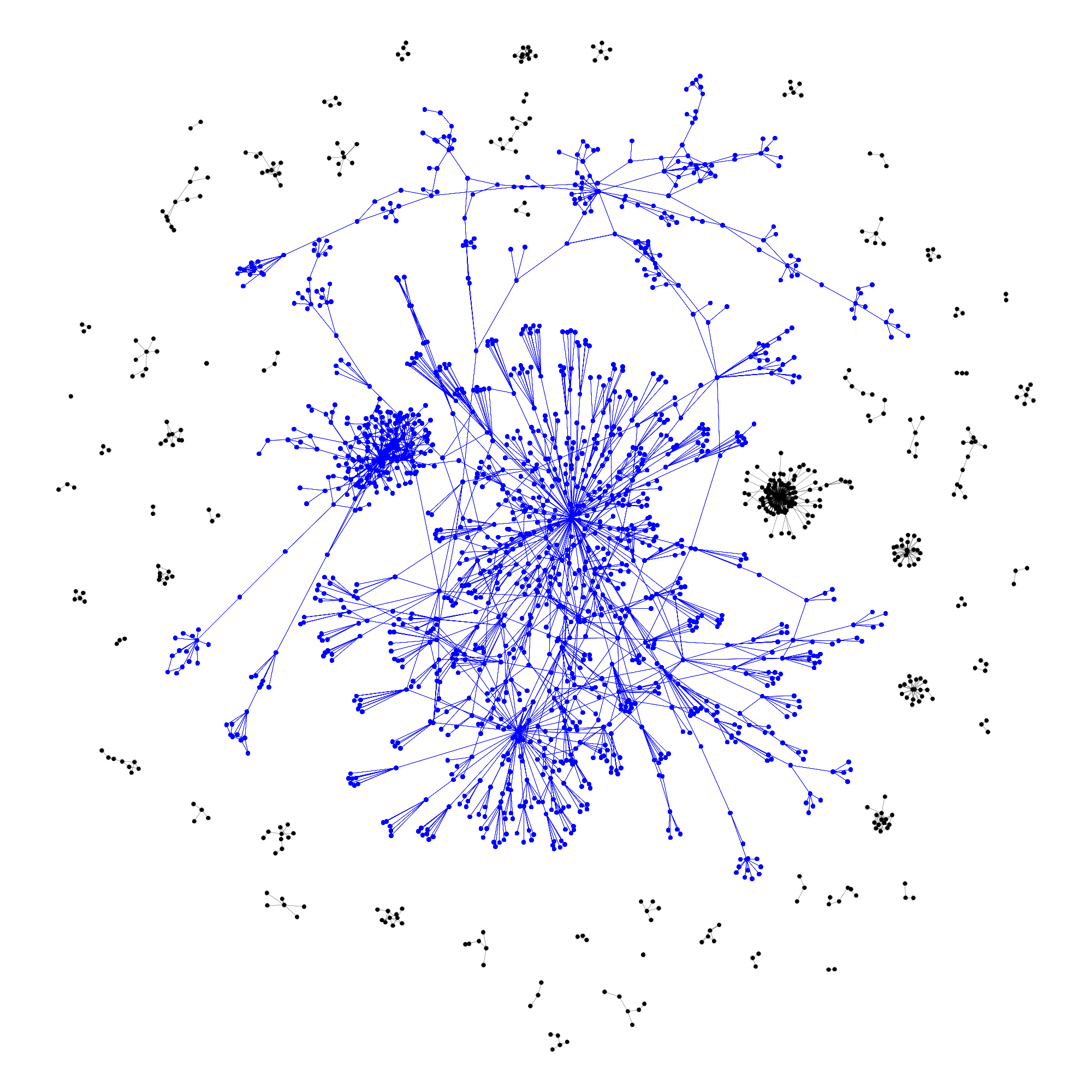}%
\caption{Visualization of the network structure of AIDA sentences with their associated publications (left), additionally augmented with some known domain identifiers (middle), and with automatically detected DBpedia connections (right). The largest connected component is shown in blue.}
\label{fig:network}
\end{figure}

As a next step, we can look into the degree to which we achieved the goal of a connected network of scientific findings. Well-connected networks allow for effectively browsing, searching, aggregating, and clustering this conceptual space of scientific knowledge.

We can first look at the structure of the network consisting of AIDA sentences and the publications they link to. Unsurprisingly, this leads to a network of many small disconnected components, as shown on the left hand side of Figure \ref{fig:network}. The network consists of 989 nodes (615 of them unique AIDA sentences, with identical sentences merged from the initial set of 650 sentences) forming 332 network components (i.e. internally connected sub-networks without connections to other components). The AIDA sentences from the Alzheimer's study form the largest component with 62 AIDA nodes (10.1\%), which are connected via the node for the respective meta-review publication.
Already before applying our automatic linking, some of the AIDA sentences came with links to domain entities: The sentences from our previous study on the automatic extraction in the biomedical domain came with formal links to gene and organism identifiers. Adding these links to the network produces a large new component, comprising of 149 AIDA sentences (24.2\%), shown in the middle part of Figure \ref{fig:network}. The network is, however, still dominated by many small disconnected components.

On this background, we can now assess the impact of our automatic DBpedia linking. It introduced additional 711 nodes to the network, representing 711 distinct DBpedia concepts. The biggest component has now grown to include almost half of all AIDA nodes (48.1\%, or 296 out of 615) and the number of components is reduced to 66. 80.1\% of the components of the initial network therefore have been merged by the added links. The network starts to show a more complex structure, as can be seen on the right hand part of Figure \ref{fig:network}. About half of the AIDA sentences, therefore, can now be found by browsing through the largest component. These network results do not form a direct proof but can be seen as an indication that a dense and useful knowledge structure starts appearing when applying such automatic linking methods.

All data, code, and results from the presented studies can be found online.\footnote{\url{https://github.com/tkuhn/aida}}

\section*{Conclusions}

AIDA sentences are designed to improve organization and communication of scientific knowledge by establishing an intuitive and general formalism to identify and link scientific claims. This formalism, in turn, can be used as a basis for further formalization in the ``upward'' as well as ``downward'' direction. Upward formalization can establish the different kinds of relationships and aggregations with formal relations expressing things like ``is more general than'' or ``follows from''. Downward formalization can take AIDA sentences as a anchor to attach partial or complete domain-level representations, for example involving relations like ``is a disease affected by gene'' or ``correlates in human populations with''.
Formal representations get complicated at the very low as well as the very high level. The AIDA approach boils down to the idea that the most practical method might be to start from the middle layer of individual statements and to grow the formalization from there in both directions, and then to see how far we get.

The preliminary results presented here indeed indicate that the approach is promising. Students confirmed that AIDA sentences summarize research findings in a useful manner. Our linking and network study moreover showed that such findings can be automatically connected to Linked Data identifiers at good accuracy and that this process leads to a dense and broad network of scientific findings.

As future work, I plan to work on providing a publishing infrastructure for such AIDA sentences, based on the concept and infrastructure of a Linked Data format called nanopublications \cite{kuhn2016decentralized}. With these techniques, AIDA sentences could be published, corrected, reviewed, linked, searched, and aggregated in a fully decentralized and open manner. Researchers could then publish their latest findings as AIDA sentences right away, and link it to the existing body of scientific findings.

\bibliography{aida}{}

\begin{thebibliography}{10}
\providecommand{\url}[1]{#1}
\csname url@samestyle\endcsname
\providecommand{\newblock}{\relax}
\providecommand{\bibinfo}[2]{#2}
\providecommand{\BIBentrySTDinterwordspacing}{\spaceskip=0pt\relax}
\providecommand{\BIBentryALTinterwordstretchfactor}{4}
\providecommand{\BIBentryALTinterwordspacing}{\spaceskip=\fontdimen2\font plus
\BIBentryALTinterwordstretchfactor\fontdimen3\font minus
  \fontdimen4\font\relax}
\providecommand{\BIBforeignlanguage}[2]{{%
\expandafter\ifx\csname l@#1\endcsname\relax
\typeout{** WARNING: IEEEtran.bst: No hyphenation pattern has been}%
\typeout{** loaded for the language `#1'. Using the pattern for}%
\typeout{** the default language instead.}%
\else
\language=\csname l@#1\endcsname
\fi
#2}}
\providecommand{\BIBdecl}{\relax}
\BIBdecl

\bibitem{bastian2010seventy}
H.~Bastian, P.~Glasziou, and I.~Chalmers, ``Seventy-five trials and eleven
  systematic reviews a day: how will we ever keep up?'' \emph{PLoS medicine},
  vol.~7, no.~9, p. e1000326, 2010.

\bibitem{nadeau2007survey}
D.~Nadeau and S.~Sekine, ``A survey of named entity recognition and
  classification,'' \emph{Lingvisticae Investigationes}, vol.~30, no.~1, pp.
  3--26, 2007.

\bibitem{wei2015overview}
C.-H. Wei, Y.~Peng, R.~Leaman, A.~P. Davis, C.~J. Mattingly, J.~Li, T.~C.
  Wiegers, and Z.~Lu, ``Overview of the {BioCreative} {V} chemical disease
  relation {(CDR)} task,'' in \emph{Proceedings of the fifth BioCreative
  challenge evaluation workshop}.\hskip 1em plus 0.5em minus 0.4em\relax
  Sevilla Spain, 2015, pp. 154--166.

\bibitem{ciccarese2011open}
P.~Ciccarese, M.~Ocana, L.~J.~G. Castro, S.~Das, and T.~Clark, ``An open
  annotation ontology for science on web 3.0,'' in \emph{Journal of biomedical
  semantics}, vol.~2, no.~2.\hskip 1em plus 0.5em minus 0.4em\relax BioMed
  Central, 2011, p.~S4.

\bibitem{kuhn2013eswc}
T.~Kuhn, P.~E. Barbano, M.~L. Nagy, and M.~Krauthammer, ``Broadening the scope
  of nanopublications,'' in \emph{Proceedings of ESWC}.\hskip 1em plus 0.5em
  minus 0.4em\relax Springer, 2013, pp. 487--501.

\bibitem{kuhn2014cl}
\BIBentryALTinterwordspacing
T.~Kuhn, ``A survey and classification of controlled natural languages,''
  \emph{Computational Linguistics}, vol.~40, no.~1, pp. 121--170, March 2014.
  [Online]. Available:
  \url{http://www.mitpressjournals.org/doi/abs/10.1162/COLI_a_00168}
\BIBentrySTDinterwordspacing

\bibitem{kuhn2017genuine}
T.~Kuhn and M.~Dumontier, ``Genuine semantic publishing,'' \emph{Data Science},
  vol.~1, no. 1--2, 2017.

\bibitem{dewaard2012formalising}
A.~De~Waard and J.~Schneider, ``Formalising uncertainty: An ontology of
  reasoning, certainty and attribution {(ORCA)},'' in \emph{Proceedings of the
  Joint 2012 International Conference on Semantic Technologies Applied to
  Biomedical Informatics and Individualized Medicine-Volume 930}.\hskip 1em
  plus 0.5em minus 0.4em\relax CEUR-WS. org, 2012, pp. 10--17.

\bibitem{chibucos2014standardized}
M.~C. Chibucos, C.~J. Mungall, R.~Balakrishnan, K.~R. Christie, R.~P. Huntley,
  O.~White, J.~A. Blake, S.~E. Lewis, and M.~Giglio, ``Standardized description
  of scientific evidence using the evidence ontology {(ECO)},''
  \emph{Database}, vol. 2014, 2014.

\bibitem{kuhn2012wole}
T.~Kuhn and M.~Krauthammer, ``Underspecified scientific claims in
  nanopublications,'' in \emph{Proceedings of the Workshop on the Web of Linked
  Entities (WoLE 2012)}.\hskip 1em plus 0.5em minus 0.4em\relax CEUR-WS, 2012,
  pp. 29--32.

\bibitem{jansen2016bnaic}
T.~Jansen and T.~Kuhn, ``Extracting core claims from scientific articles,'' in
  \emph{Benelux Conference on Artificial Intelligence}.\hskip 1em plus 0.5em
  minus 0.4em\relax Springer, 2016, pp. 32--46.

\bibitem{ogden1930basic}
C.~K. Ogden, \emph{{B}asic {E}nglish: a general introduction with rules and
  grammar}.\hskip 1em plus 0.5em minus 0.4em\relax London: Paul Treber \& Co.,
  1930.

\bibitem{kuhn2006improving}
T.~Kuhn, L.~Royer, N.~E. Fuchs, and M.~Schroeder, ``Improving text mining with
  controlled natural language: A case study for protein interactions,'' in
  \emph{International Workshop on Data Integration in the Life Sciences}.\hskip
  1em plus 0.5em minus 0.4em\relax Springer, 2006, pp. 66--81.

\bibitem{hallett2007coli}
C.~Hallett, D.~Scott, and R.~Power, ``Composing questions through conceptual
  authoring,'' \emph{Computational Linguistics}, vol.~33, no.~1, pp. 105--133,
  2007.

\bibitem{kuhn2012corpora}
T.~Kuhn and S.~H\"ofler, ``{C}oral: Corpus access in controlled language,''
  \emph{Corpora}, vol.~7, no.~2, pp. 187--206, 2012.

\bibitem{bizer2009linked}
C.~Bizer, T.~Heath, and T.~Berners-Lee, ``Linked data-the story so far,''
  \emph{International journal on semantic web and information systems}, vol.~5,
  no.~3, pp. 1--22, 2009.

\bibitem{mendes2011dbpedia}
P.~N. Mendes, M.~Jakob, A.~Garc{\'\i}a-Silva, and C.~Bizer, ``{DBpedia}
  spotlight: shedding light on the web of documents,'' in \emph{Proceedings of
  the 7th international conference on semantic systems}.\hskip 1em plus 0.5em
  minus 0.4em\relax ACM, 2011, pp. 1--8.

\bibitem{auer2007dbpedia}
S.~Auer, C.~Bizer, G.~Kobilarov, J.~Lehmann, R.~Cyganiak, and Z.~Ives,
  ``{Dbpedia}: A nucleus for a web of open data,'' in \emph{The semantic
  web}.\hskip 1em plus 0.5em minus 0.4em\relax Springer, 2007, pp. 722--735.

\bibitem{kuhn2016decentralized}
T.~Kuhn, C.~Chichester, M.~Krauthammer, N.~Queralt-Rosinach, R.~Verborgh,
  G.~Giannakopoulos, A.-C.~N. Ngomo, R.~Viglianti, and M.~Dumontier,
  ``Decentralized provenance-aware publishing with nanopublications,''
  \emph{PeerJ Computer Science}, vol.~2, p. e78, 2016.

\end{thebibliography}
\bibliographystyle{IEEEtran}

\end{document}